%% file: main.tex
\documentclass[11pt,prd,reprint,twocolumn,notitlepage,nofootinbib]{revtex4-1}

\usepackage{float}
\usepackage{amsmath,bm}
\usepackage{amsfonts}
\usepackage{graphicx, adjustbox}
\usepackage{multirow}
\usepackage{aas_macros}
\usepackage{dblfloatfix}  
\usepackage[title]{appendix}
\usepackage[usenames]{xcolor}
\usepackage{hyperref}

\definecolor{mpl_blue}{HTML}{1F77B4}
\definecolor{mpl_orange}{HTML}{FF7F0E}
\definecolor{mpl_green}{HTML}{2CA02C}
\definecolor{mpl_red}{HTML}{D62728}

\newcommand{\libstempo}{\texttt{libstempo}}
\newcommand{\enterprise}{\texttt{enterprise}}

\newcommand{\entext}{\texttt{enterprise\_extensions}}
\newcommand{\ptmcmc}{\texttt{PTMCMCSampler}}

\date{\today}

\begin{document}

\title{Generalized optimal statistic for characterizing multiple correlated signals in pulsar timing arrays}

\author{Shashwat C.~Sardesai}
\affiliation{Center for Gravitation, Cosmology and Astrophysics, University of Wisconsin--Milwaukee, P.O. Box 413, Milwaukee WI, 53201, USA}

\author{Sarah J.~Vigeland}
\affiliation{Center for Gravitation, Cosmology and Astrophysics, University of Wisconsin--Milwaukee, P.O. Box 413, Milwaukee WI, 53201, USA}

\author{Kyle A.~Gersbach}
\affiliation{Department of Physics and Astronomy, Vanderbilt University, 2301 Vanderbilt Place, Nashville, Tennessee 37235, USA}

\author{Stephen R.~Taylor}
\affiliation{Department of Physics and Astronomy, Vanderbilt University, 2301 Vanderbilt Place, Nashville, Tennessee 37235, USA}

\begin{abstract}
Pulsar timing arrays are sensitive to low-frequency gravitational waves (GWs), 
including the low-frequency stochastic gravitational wave background (GWB), 
which induces correlated changes in millisecond pulsars' timing residuals 
described by the Hellings-Downs curve. 
Some sources of noise can also induce correlated changes in pulsar timing residuals, 
albeit with different correlation signatures. 
A spatial correlation that differs from Hellings-Downs could also be indicative of non-Einsteinian GW polarizations. 
It is therefore crucial that we be able to characterize the spatial correlation 
in order to distinguish between the GWB and sources of noise. 
The optimal statistic (OS) is a frequentist estimator for the amplitude and significance of a spatially-correlated signal 
in PTA data, and it is widely used to search for the GWB. 
However, the OS cannot perfectly distinguish between different spatial correlations. 
In this paper, we introduce the multiple component optimal statistic (MCOS): a generalization of the OS that allows for multiple correlations to be simultaneously fit to the data. 
We use simulated data to show that this method more accurately recovers injected spatially correlated signals, 
and in particular eliminates the problem of overestimating the amplitude of correlations that are not present in the data. 
We also demonstrate that this method can be used to recover multiple correlated signals.
\end{abstract}

\maketitle

\section{Introduction}

When massive bodies such as black holes orbit each other, they radiate away energy in the form of gravitational waves (GWs), causing their orbit to shrink until they coalesce. Supermassive black hole binaries (SMBHBs), which form in galaxy mergers, emit low-frequency GWs over many orders of magnitude in frequency 
as they go from subparsec separations to coalescence. 
The incoherent superposition of GWs from a cosmological population of SMBHBs 
produces a stochastic GW background (GWB) in the nanohertz frequency range 
($\sim 10^{-9} - 10^{-7} \; \mathrm{Hz}$) \cite{2013CQGra..30x4009S, 2004ApJ...611..623S, 2019A&ARv..27....5B}.

Pulsars are rapidly spinning neutron stars that emit radio waves that can be observed by radio telescopes. Pulsars have very stable spin rates, 
which makes them ideal for detecting long-period gravitational waves 
\cite{1978SvA....22...36S, 1979ApJ...234.1100D, 1990ApJ...361..300F}. 
In pulsar timing arrays, pulsars essentially act as astronomical clocks, 
allowing us to measure small fluctuations in spacetime over years to decades. 
There are currently four regional PTA experiments: 
the North American Nanohertz Observatory for Gravitational Waves (NANOGrav) \cite{brazier+2019}, 
the European Pulsar Timing Array (EPTA) \cite{2016MNRAS.458.3341D}, 
the Parkes Pulsar Timing Array (PPTA) \cite{2016MNRAS.455.1751R}, 
and the Indian Pulsar Timing Array (InPTA) \cite{2022PASA...39...53T}. 
All of these groups collaborate and share data under the umbrella of the 
International Pulsar Timing Array (IPTA) \cite{pdd+2019}.

The presence of GWs induces correlated changes in the pulse times of arrival, 
with the correlation between two different pulsars depending on their angular separation 
on the sky according to the Hellings-Downs (HD) curve \cite{1983ApJ...265L..39H}. 
This characteristic correlation allows us to distinguish between the GWB and 
other astrophysical and terrestrial effects that could affect the pulse times of arrival of many pulsars, such as 
the solar wind, instrumentation errors, a clock error, 
ephemeris errors, 
etc.~\cite{2012MNRAS.427.2780H,2016MNRAS.455.4339T,2020MNRAS.491.5951H}. 
A correlated signal with a correlation pattern 
that differs from the HD curve could also indicate the presence of 
non-Einsteinian GW polarizations \cite{Lee_2008,2012PhRvD..85h2001C,Gair_2015,Cornish_2018}.

It is therefore crucial to be able to distinguish 
between different cross-correlations. 
One method for doing this is using the OS, 
a frequentist estimator of the amplitude and significance of a correlated 
stochastic process, where the correlation function of interest 
is determined by specifying the overlap reduction function (ORF) 
\cite{2009PhRvD..79h4030A, 2013ApJ...762...94D, 2015PhRvD..91d4048C}. 
One limitation of the OS is that for any real PTA, 
it is not possible to perfectly distinguish between different ORFs. 
In general, the ORFs for different 
cross-correlations are not orthogonal, 
with the overlap between them 
depending on the number of pulsars and their sky locations \cite{2018PhRvD..98d4003V,2021ApJ...923L..22A}. 
In this paper, 
we address this issue by introducing a modified version of the 
OS that allows us to 
simultaneously search for multiple correlated OS signals (MCOS). As we show in this paper, 
this allows us to more accurately characterize the presence 
of correlated signals in PTA data.

This paper is organized as follows. 
In Sec.~\ref{sec:os}, we discuss the OS 
and derive a modified version that can simultaneously fit 
multiple correlation functions. 
In Sec.~\ref{sec:methods}, we describe the methods, signal models, 
and software used to generate and analyze simulated PTA data. 
We generated three types of simulated data sets: 
one containing a GWB (i.e., a common-spectrum process 
with HD correlations), another containing a common-spectrum process 
with GW-like monopole correlations, and finally a model with both a GWB, and GW-like monopole signal injected.
In Sec.~\ref{sec:results} we use the standard OS and 
our MCOS to analyze the simulated data sets. 
We find that simultaneously fitting multiple correlations 
using the MCOS prevents overestimating 
the presence of correlations that are not present, 
while not significantly affecting parameter estimation 
of correlated signals. 
We also demonstrate that the MCOS allows us 
to recover multiple correlated signals from the data. 
We summarize the paper and discuss future work in Sec.~\ref{sec:conclusions}.

\section{Optimal Statistic}
\label{sec:os}

The OS is a frequentist estimator of the amplitude and significance of the stochastic background. 
It only considers the cross-correlations between two different pulsars and not auto-correlation terms of individual pulsars. 
The OS is very fast to compute compared to performing an equivalent Bayesian analysis; 
however, the OS can be biased due to covariance between individual pulsar red noise and a common stochastic process. 
The noise-marginalized OS is a hybrid Bayesian-frequentist method that 
marginalizes over the pulsars' intrinsic red noise \cite{2018PhRvD..98d4003V}. 
It has been shown that this method more accurately estimates the amplitude of the GWB 
in the presence of intrinsic pulsar red noise compared to computing the OS with fixed pulsar red noise parameters.

One method to derive the OS
is to consider fitting the cross-correlations between pulsars to an ORF $\Gamma_{ab}$ with amplitude $A^2$ \cite{2013ApJ...762...94D, 2015PhRvD..91d4048C}. Let $\mathbf{r}_a$ be the residuals for pulsar $a$. The cross-correlations between pulsars $a$ and $b$, $\rho_{ab}$, and their associated uncertainties, $\sigma_{ab}$, can be written as
\begin{eqnarray}
    \rho_{ab} &=& \mathcal{N}_{ab} \mathbf{r^{T}_{a} P^{-1}_{a} \Bar{S}_{ab} P^{-1}_{b} r_{b}} \,, \label{eq:rho_ab} \\
    \sigma_{ab}^2 &=& \mathcal{N}_{ab} = (\mathrm{tr}[\mathbf{P^{-1}_{a} \Bar{S}_{ab} P^{-1}_{b} \Bar{S}_{ba}} ])^{-1} \,, \label{eq:sig_ab}
\end{eqnarray}
\noindent where $\mathbf{P}_a$ is the auto-covariance matrix and contain contributions from white noise, intrinsic red noise, and the common stochastic process red noise from pulsar $a$; and $\mathbf{S}_{ab} = A^2 \, \Gamma_{ab} \mathbf{\Bar{S}}_{ab}$ is the cross-covariance matrix and only considers contributions from the common stochastic process between pulsars $a$ and $b$.

We can fit the measurement cross-correlations to an ORF. The chi-squared is given by
\begin{equation} \label{eq:CHI2}
	\chi^2 = \sum_{ab,a<b} \left( \frac{\rho_{ab} - A^2 \, \Gamma_{ab}}{\sigma_{ab}} \right)^2 \,,
\end{equation}
where $\sigma_{ab}$ are the uncertainties in the cross-correlations and contain contributions from both the common process and intrinsic noise, as shown in \cite{2015PhRvD..91d4048C}. 
The OS is the value of $A^2$ that minimizes the chi-squared:
\begin{equation} \label{eq:os_A2}
	\hat{A}^2 = \frac{\sum_{ab} (\rho_{ab} \Gamma_{ab}/\sigma_{ab}^2)}{\sum_{ab} (\Gamma_{ab}^2/\sigma_{ab}^2)} \,.
\end{equation}
The uncertainty in $\hat{A}^2$ can be found through simple propagation of errors:
\begin{equation} \label{eq:os_err}
	\sigma_{\hat{A}^2}^2 = \left[ \sum_{ab} \left(\frac{\Gamma_{ab}^2}{\sigma_{ab}^2}\right) \right]^{-1} \,.
\end{equation}
The corresponding signal-to-noise ratio (S/N) is
\begin{equation} \label{eq:10}
    \mathrm{S}/\mathrm{N} = \frac{\hat{A}^2}{\sigma_{\hat{A}^2}} \,.
\end{equation}

The OS as derived above assumes the presence of only one spatially correlated process in our data. 
It is possible that PTA data could contain multiple spatially correlated processes, 
e.g., a GWB and a common correlated source of noise.  
As discussed in \citet{2016MNRAS.455.4339T}, some noise sources can produce spatially correlated signals -- 
a clock error appears in PTA data as a common process with monopolar correlations 
between different pulsars, while an ephemeris error produces dipolar correlations.

If more than one spatially-correlated process is present, the chi-squared becomes
\begin{equation} \label{eq:CHI2 MCOS}
	\chi^2 = \sum_{ab} \left( \frac{\rho_{ab} - \sum_\alpha A_\alpha^2 \, \Gamma^\alpha_{ab}}{\sigma_{ab}} \right)^2 \,,
\end{equation}
where $\alpha$ indexes the different spatially-correlated signals. Minimizing the chi-squared with respect to the amplitude squared value $A_\alpha^2$ gives
\begin{equation} \label{eq:CAB}
	C^\beta - \sum_\alpha \hat{A}_\alpha^2 B^{\alpha\beta} = 0 \,,
\end{equation}
where
\begin{eqnarray}
	C^\beta &\equiv& \sum_{ab} \frac{\rho_{ab} \Gamma_{ab}^\beta}{\sigma_{ab}^2} \,, \label{eq:MCOS_C} \\
	B^{\alpha\beta} &\equiv& \sum_{ab} \frac{\Gamma_{ab}^\alpha \Gamma_{ab}^\beta}{\sigma_{ab}^2} \,. \label{eq:MCOS_B}
\end{eqnarray}
Therefore, $\hat{A}^2_\alpha$ is given by
\begin{equation} \label{eq:OS_MCOS_A2}
	\hat{A}^2_\alpha = \sum_\beta B_{\alpha\beta} C^\beta \,,
\end{equation}
where $B_{\alpha\beta}$ is the inverse of $B^{\alpha\beta}$. 
This is a more general form of Eq.~\eqref{eq:os_A2}: 
when we generalize to multiple ORFs, 
the numerator of Eq.~\eqref{eq:os_A2} becomes $C^\beta$, 
while the denominator becomes $B_{\alpha\beta}$. 
The uncertainty in $\hat{A}^2_\alpha$ is
\begin{equation} \label{eq:OS_MCOS_err}
    \sigma_{\hat{A}^2_\alpha}^2 = B_{\alpha\alpha} \,.
\end{equation}

\section{Methods and Software}
\label{sec:methods}

In this section we describe how to construct the pulsar timing model, including the intrinsic and common red noise, 
and discuss how to retrieve the injected signals. 
We based our simulated data sets on the NANOGrav 12.5-year data set \cite{2021ApJS..252....4A}, 
with the pulsar positions, observing spans, cadences, and white and red noise all taken from that data set. 
In addition, we inject two different kinds of stochastic backgrounds: 
one with HD correlations (i.e., a GWB), and one with GW-like monopolar correlations. 
The methods used are similar to those used in the NANOGrav 12.5-year GWB paper \cite{2020ApJ...905L..34A}: 
below we briefly describe them, highlighting how our methods differ from those in \cite{2020ApJ...905L..34A}.

\subsection{PTA model}

We model the timing residuals for a pulsar $\delta \bold{t}$ as
    \begin{equation} \label{eq:PTA model}
       \delta \bold{t} = M \epsilon + F \bold{a} + \bold{n} \,,
    \end{equation}
\noindent where $M \epsilon$ describes \textbf{linear} perturbations to the timing model, 
$F \bold{a}$ describes red noise, including both individual pulsar red noise 
and a common stochastic process affecting all of the pulsars, 
and $\bold{n}$ is the uncorrelated white noise. 
We model the red noise using a Fourier series with frequencies $f = 1/T, 2/T, \ldots$, 
where $T$ is the span of the observations, 
and use a power-law model for the power spectral density 
\begin{equation}
    P(f) = \frac{A^2}{12 \pi^2} \left(\frac{f}{f_\mathrm{yr}}\right)^{-\gamma} \mathrm{yr}^3 \,,
    \label{eq:rn_psd}
\end{equation}
where $A$ is the amplitude, 
defined at a reference frequency of $f_\mathrm{yr} = 1/(1 \; \mathrm{year})$, 
and $\gamma$ is the spectral index. 

\subsection{Common stochastic process}

In addition to pulsar red noise and white noise, 
we also include the presence of a common stochastic process. 
For all types of stochastic backgrounds, 
we model the power spectrum as a power law as described in Eq.~\eqref{eq:rn_psd} with $A_\mathrm{gw} = 2 \times 10^{-15}$ and $\gamma = 13/3$. 
The value of $\gamma$ corresponds to that for a GWB made from GWs from 
circular SMBBHs evolving only due to GW emission \cite{2001astro.ph..8028P}. 

We model two different types of stochastic process: 
one with HD correlations, as is expected for a GWB;  
one with spatial correlations described by a GW-like monopole (GWMO). 
The ORF for HD correlations is \cite{1983ApJ...265L..39H, 2015AmJPh..83..635J}
\begin{eqnarray}
    \Gamma^{\mathrm{HD}}_{ab}(\xi) &=& \frac{\delta_{ab}}{2} + \frac{1}{2} - \frac{(1-\cos\xi)}{4} \left[\frac{1}{2} \right. \nonumber \\
        && \left. - 3 \ln \left(\frac{1-\cos\xi}{2}\right)\right] \,, \label{eq:ORF HD}
\end{eqnarray}
where $\Gamma^{\mathrm{HD}}_{ab}$ is the HD ORF for pulsars with indices $a$ and $b$, and 
$\xi$ is the angular separation between them. 
Note that the maximum correlation between two different pulsars (i.e., $a \neq b$) is $1/2$: 
this is because the GWB induces an Earth term and a pulsar term in each pulsar's residuals, 
and only the Earth terms are correlated. 
The GWMO was introduced in \cite{2021ApJ...923L..22A} and 
the ORF takes the form
\begin{equation}
    \Gamma^{\mathrm{GWMO}}_{ab}(\xi) = \frac{1}{2} + \frac{\delta_{ab}}{2} \,. \label{eq:ORF GWMO}
\end{equation}
The cross-correlation for two different pulsars is $1/2$ rather than $1$ 
because we are assuming the signal has two components: an ``Earth term'' that is correlated 
and a ``pulsar term'' that is uncorrelated (hence why it is described as ``GW-like''). 
The GWMO does not correspond to any physical process, 
but it is similar to the ORF 
produced by a scalar-tensor mode.

\subsection{Simulated data sets}

We use \libstempo~\cite{vallisneri_libstempo_2020} to generate our simulated data sets. 
The simulated data sets are based on the NANOGrav 12.5-year data set \cite{2021ApJS..252....4A}. 
We create two types of simulated data sets. 
One contains a common stochastic process with HD correlations. 
The second contains two commmon stochastic processes: 
one with HD correlations and another with GWMO correlations. 
For both, 
we use the linearized timing models for the pulsars in the 12.5-year data set, 
as well as the dates and times of each observation. We first idealize our pulsar timing residuals and then add uncorrelated white noise 
equal to the measured TOA uncertainties. 
We add pulsar intrinsic red noise, with the maximum likelihood values taken from a Bayesian run of the NANOGrav 12.5-year data set with the values given (see Appendix \ref{app:pulsar_noise}). 

\subsection{Optimal statistic calculation}

In order to compute the OS, we must specify values for the pulsars' intrinsic white noise and red noise. 
The white noise parameters are the maximum-likelihood values from Bayesian noise analyses where each pulsar is run individually. 
The red noise parameters come from a Bayesian analysis of all the pulsars that simultaneously fits for pulsar intrinsic red noise 
and a common uncorrelated stochastic process. The white noise parameters are not searched over in the Bayesian analysis 
and are instead fixed to the maximum-likelihood values. 
We use uniform priors for the red noise parameters, 
$\log_{10}A \in [-20,-11]$ and $\gamma \in [0, 7]$, 
and the common process amplitude, 
$\log_{10} A_\mathrm{gw} \in [-18, -11]$, while the common process $\gamma_{\mathrm{gw}}$ is fixed to 13/3.
We use \enterprise~\cite{enterprise} and \entext~\cite{ent_ext} to implement the models and compute the likelihood, 
and we use Markov Chain Monte Carlo (MCMC) methods to obtain samples from the posterior 
as implemented in \ptmcmc~\cite{ptmcmc}.

We then compute the OS, 
using the results of the Bayesian analysis 
to determine the pulsar intrinsic red noise 
and common uncorrelated stochastic process. 
There are two ways of doing this. 
The \textit{fixed-noise} version computes the OS at a single noise realization using the red noise values that maximize the likelihood of the Bayesian analysis. 
The \textit{noise-marginalized} version pulls red noise values from the posteriors obtained by the Bayesian analysis in order to compute the OS at many different noise realizations. 
This results in distributions for $\hat{A}^2$ and $\sigma$. 

\begin{figure}[ht]
    \centering
    \includegraphics[width=\columnwidth]{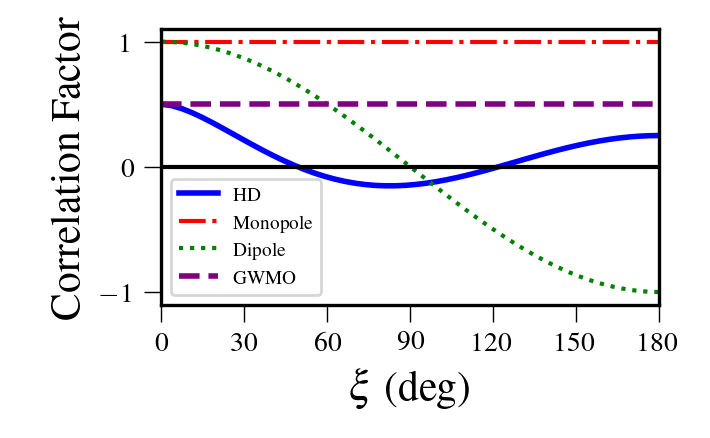}
    \caption{Comparison between HD (blue), monopole (red), dipole (green), and GWMO (purple) ORFs as a function of angular separation between pulsars. The solid black line indicates a zero amplitude correlation.
    Note that a GWMO has the same functional form as the standard monopole, 
    but a different normalizaton.
    }
    \label{fig:ORFs}
\end{figure}

We can compute the OS for different correlations by using different ORFs. 
In addition to the HD and GWMO ORFs given in Eqs.~\eqref{eq:ORF HD} and \eqref{eq:ORF GWMO}, 
we also consider a standard dipole ORF and monopole ORF:
\begin{eqnarray} \label{eq:ORFs}
    \Gamma^{\mathrm{dipole}}_{ab}(\xi) &=& \cos\xi \,, \\ \label{eq:ORF DI}
    \Gamma^{\mathrm{monopole}}_{ab}(\xi) &=& 1 \,. \label{eq:ORF MO}
\end{eqnarray}
In Figure~\ref{fig:ORFs}, we plot all four ORFs used in this paper. Note that the monopole and GWMO are both constant as a function of angular separation, 
but the monopole as a value of $1$ while the GWMO as a value of $1/2$. This means that any signal that fits one will also fit the other, 
but the inferred amplitudes will differ because the two ORFs have different normalizations. 

We can measure the overlap between different ORFs by computing the ``\textbf{unweighted} match statistic'' \cite{2016PhRvD..93j4047C},
\begin{equation} \label{eq:orf_match}
    \Bar{M} = \frac{\sum_{a,b\neq a} \Gamma_{ab} \Gamma'_{ab}}{\sqrt{(\sum_{a,b\neq a} \Gamma_{ab} \Gamma_{ab}) (\sum_{a,b\neq a} \Gamma'_{ab} \Gamma'_{ab})}} \,,
\end{equation}
where $\Gamma$ and $\Gamma'$ are two different ORFs. 
This overlap depends on the number of pulsars in the PTA and their sky locations. 
Table \ref{tab:matchstat} lists the match statistics for the four ORFs considered in this paper for our simulated PTA, which is based on the NANOGrav 12.5-year data set.

\input{table_orfs}

\section{Results}
\label{sec:results}

Here we present the results of analyzing our simulated data sets 
using the standard OS calculation and the modified MCOS. 
We show two types of simulations: 
one with an injected GWB (i.e., 
a common stochastic process with HD correlations), and one with both HD and GWMO correlations. 
We present the recovered OS values and 
corresponding signal-to-noise ratios (S/N) 
for different spatial correlations, 
and compare them to the injected values. We note the fraction of realizations with $\mathrm{S/N} \geq$ 3 in Appendix \ref{app:tables}.

We also look at how accurately the OS and MCOS recover the amplitude of the correlated process. If a non-negligible correlated signal is present in the data, then the expressions for $\hat{A}^2$ and $\sigma_{\hat{A}^2}$ given in Eqs.~\eqref{eq:os_A2}, \eqref{eq:os_err}, \eqref{eq:OS_MCOS_A2}, and \eqref{eq:OS_MCOS_err} 
must be revised to include interpulsar pair correlation \citep{PhysRevD.108.043026,2023arXiv230616223J}. 
For more details, see App.~\ref{app:p-p}. 
We use the recovered $\hat{A}^2$ and $\sigma_{\hat{A}^2}$ 
to compute the percentile of the injected value, 
assuming a Gaussian distribution\footnote{As shown in \cite{Hazboun:2023tiq}, the optimal statistic actually follows a generalized chi-squared distribution and not a Gaussian distribution. In this work, we approximate the distribution as a Gaussian, owing to the computational expense of constructing the generalized chi-squared distribution for each simulation. This is a reasonably good approximation, and has been used in previous work \citep{2018PhRvD..98d4003V, 2023arXiv230616223J}.}, 
and compare the fraction of simulations for

Finally, we look at how much the data preferred models 
with multiple correlated signals 
using the Akaike Information Criterion (AIC) \cite{Akaike1998}, 
which is a frequentist analog for the Bayes factor. It is given by
\begin{equation} \label{eq:AIC}
   \mathrm{AIC} = 2k + \chi^2  \,,
\end{equation}
\noindent where $k$ is the number of model parameters (i.e., the number of correlated signal amplitudes, or the number of ORFs), 
and the chi-squared given by Eq.~\eqref{eq:CHI2} for a single correlated signal 
and Eq.~\eqref{eq:CHI2 MCOS} for multiple correlated signals
The factor of $2k$ acts as an Occam's penalty for adding more parameters to a model. 
We compute $\chi^2$ using the maximum-likelihood red noise parameters. 
The relative probability of two models is then given by
\begin{equation} \label{eq:AIC comparison}
    p(\mathrm{AIC}) = e^\frac{(\mathrm{AIC} - \mathrm{AIC_{min}})}{2} \,,
\end{equation}
\noindent where $\mathrm{AIC_{min}}$ represents the minimum AIC value of all the different models. The model with the minimum AIC will thus be the most preferred model, with a p(AIC) = 1. Some of the models are equally preferred, with the difference in their AIC $\sim$ 0.01, so we set a threshold value of p(AIC) $\geq$ 0.99 to account for these cases. These results are also presented in Appendix \ref{app:tables}.

\subsection{HD simulation}

\begin{figure}[t]
    \centering
    \includegraphics[width=\columnwidth]{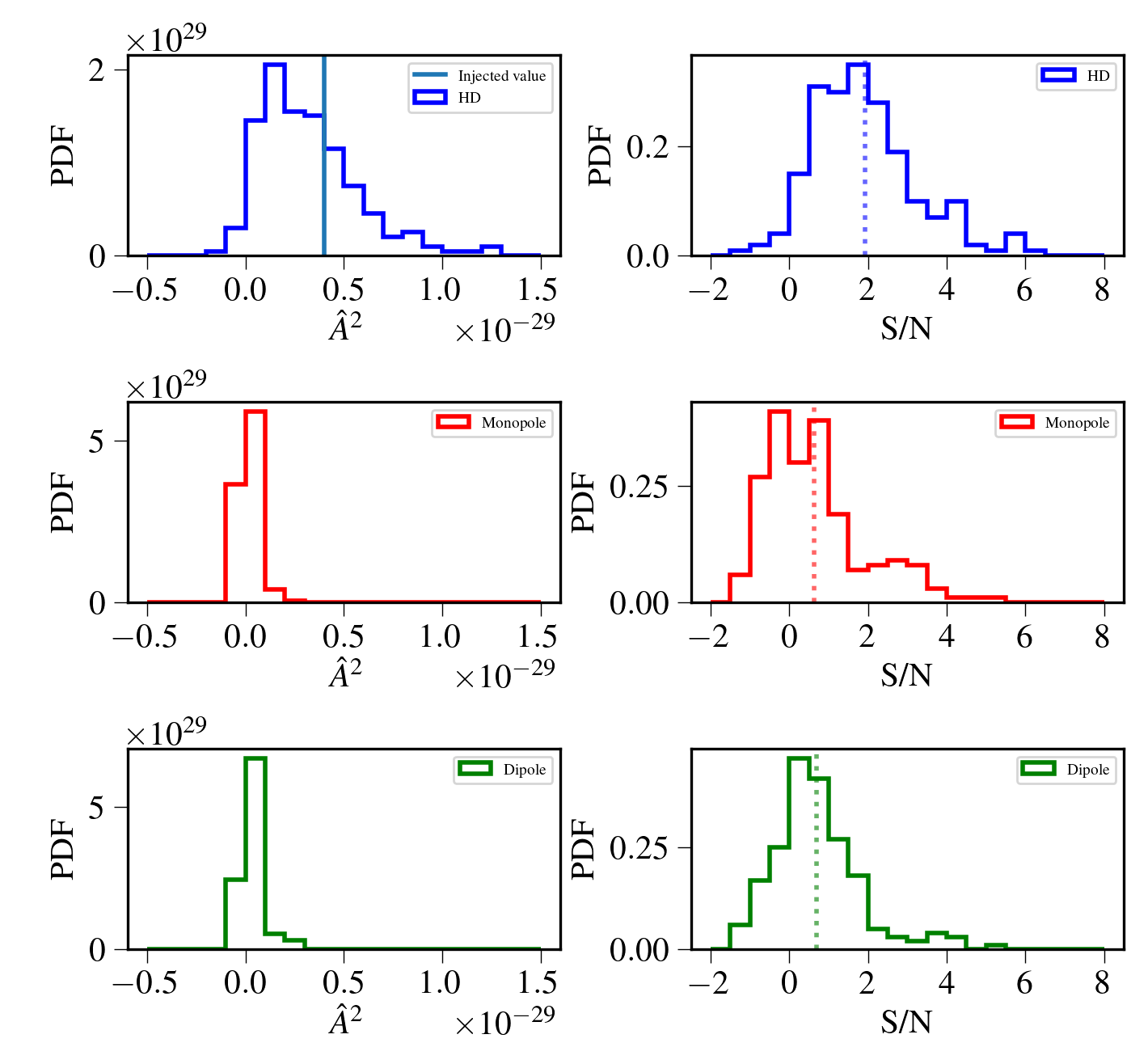}
    \vspace{-0.2in}
    \caption{Noise-marginalized OS analysis of 200 simulations
    with an injected HD. 
    We show results using HD (blue), monopole (red), and dipole (green) ORFs. The left hand columns show the distributions of $\hat{A}^2$, whereas the right hand columns show the S/N histograms.
    The vertical blue line in the top left plot indicates the injected value of $A^2_\mathrm{HD} = 4 \times 10^{-30}$. The colored dashed lines in the S/N plots are the means of the respective ORFs. We find a mean $\hat{A}^2 > 0$ and mean S/N $>0$ for monopole and dipole correlations even though neither were injected in the data.}
    \label{fig:HD_os_indv}
\end{figure}

We created 200 simulations with an injected HD 

which we analyzed using the traditional OS and the MCOS. 
A complete summary of the results can be found in Table~\ref{tab:hd_sim_results} in Appendix~\ref{app:tables}. 
When we search for one spatial correlation at a time, 
we find a S/N $>3$ for HD correlations in 17\% of the simulations, 
but we also find S/N $>3$ for monopole and dipole correlations 
in 6\% and 5\% of the simulations, respectively, even though there are no monopole-correlated or dipole-correlated signals present in the data sets. 
In Figure~\ref{fig:HD_os_indv}, 
we show histograms for the recovered distributions of $\hat{A}^2$ 
and the S/N for HD, monopole, and dipole correlations. 
On average, we recover higher S/N of HD correlations 
than for monopole or dipole correlations; 
however, we recover S/N $>0$ 
for monopole and dipole correlations in more than half of the simulations, as shown by the dotted lines in Fig. \ref{fig:HD_os_indv}.

\begin{figure}[t]
    \centering
    \includegraphics[width=\columnwidth]{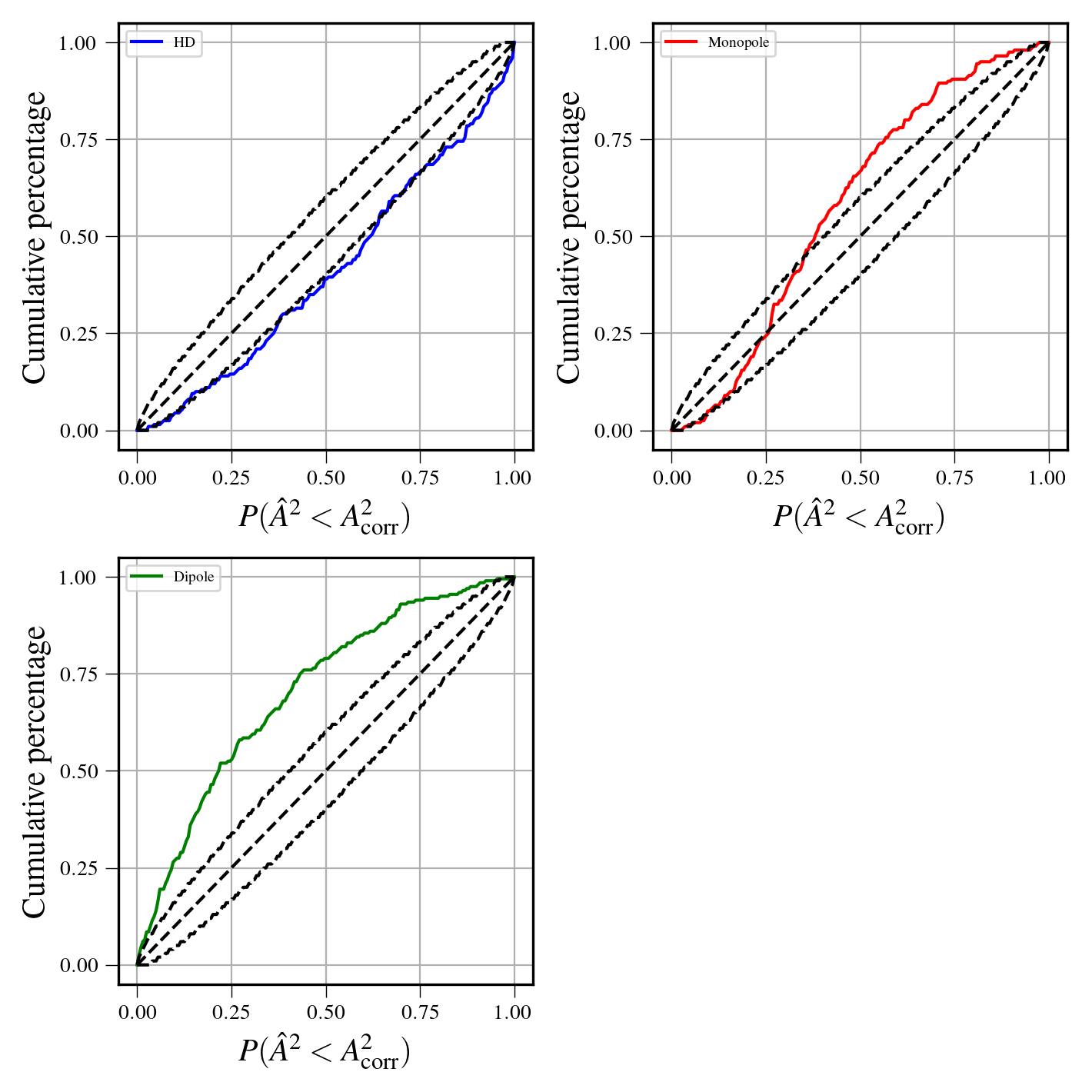}
    \vspace{-0.2in}
    \caption{$p-p$ plot for the HD injected simulations with single component OS. The horizontal axis is the percentile of each simulation to the injected value. The vertical axis represents the cumulative fraction of simulations that are recovered at that particular percentile. The models shown here are HD (top left), monopole (top right), and dipole (bottom left). If the parameters are recovered accurately, the two quantities would be equal (black dashed line). The curved black lines indicate the 95\% confidence intervals of the p-p plots. We find that the amplitude of HD correlations is underestimated, while the amplitude of monopole and dipole correlations (which are not present) are significantly overestimated.}
    \label{fig:p-p hd diagram}
\end{figure}

Figure~\ref{fig:p-p hd diagram} shows $p-p$ plots 
for the recovered single component OS $\hat{A}^2$ 
using HD, monopole, and dipole correlations.
For each simulation, we use the marginalized chain values to calculate $\hat{A}^2$ and $\sigma_{\hat{A}^2}$, taking pulsar pair cross covariance into consideration, and then compute the percentile of the injected value for that type of correlation.

We then plot the cumulative fraction of realizations for which the injected value is at that percentile. 
If the parameters are accurately recovered, 
we would expect the cumulative fraction of realizations to be equal to the cumulative percentile. 
If it is not, that indicates the parameter is being overestimated (over the central dashed black line) or underestimated (under the central dashed black line). 
We find that the amplitude of HD correlations is underestimated, 
while the amplitudes of monopole and dipole correlations (which were not injected) are overestimated. 
The underestimation of the amplitude of HD correlations is likely due to two factors: 
first, there is covariance between 
the pulsars' intrinsic red noise 
and common red noise, and second, 
the simulations inject the intrinsic red noise 
and common red noise over more frequency components than we use to recover them.

\begin{figure}[t]
    \centering
    \includegraphics[width=0.4\textwidth]{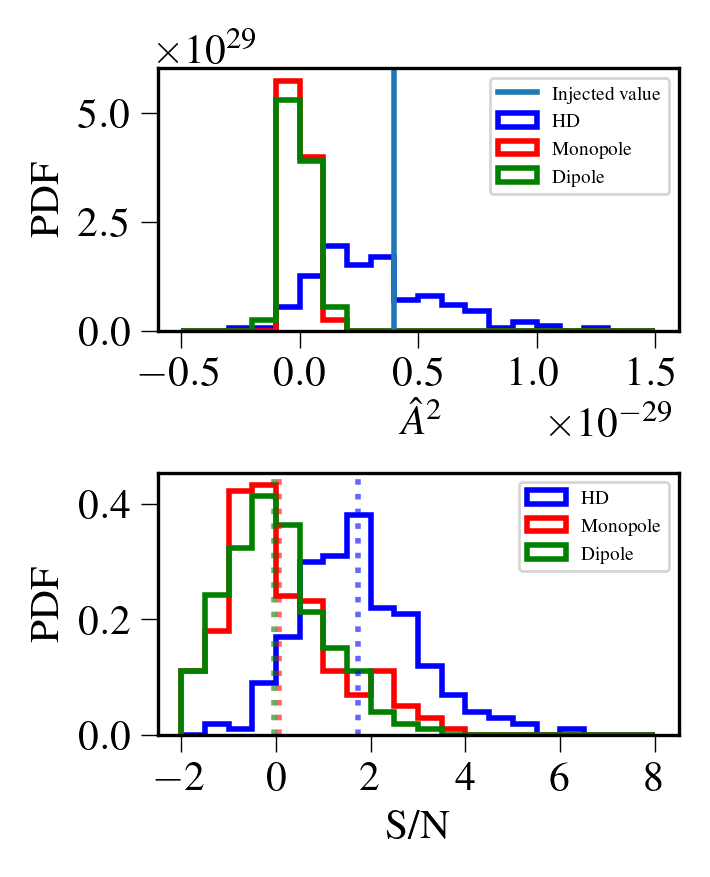}
    \vspace{-0.2in}
    \caption{MCOS analysis of 200 simulations for HD simulations for the HD+monopole+dipole ORFs. The $\bold{top \ figure}$  contains histograms of $\hat{A}^2$ and the $\bold{bottom \ figure}$ contains histograms for the S/N. The solid light blue line is the injected value of $A^2_{\mathrm{HD}} = 4 \times 10^{-30}$. The colored dashed lines in the S/N plots are the means of the respective ORFs. The recovered amplitudes and S/N for monopole and dipole correlations are centered around zero, unlike in Fig.~\ref{fig:HD_os_indv}.}
    \label{fig:HD_os_hd_mo}
\end{figure}

\begin{figure}[t]
    \centering    \includegraphics[width=0.75\columnwidth]{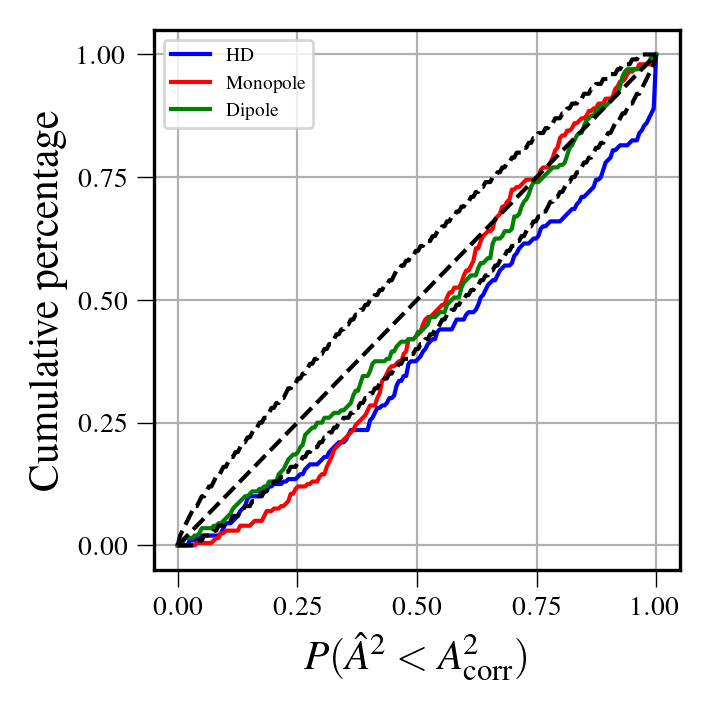}
    \vspace{-0.2in}
    \caption{$p-p$ plot for the MCOS HD injected simulations. The amplitudes shown here are recovered using a model that simultaneously fits for HD, monopole, and dipole correlations. The horizontal axis is the recovered percentile of each simulation to the injected value. The vertical axis represents the cumulative fraction of realizations that recover the particular percentile. The central black dashed line indicates accurate recovery, and the upper and lower dashed black lines indicate the 95\% confidence interval. We find that the amplitudes of monopole and dipole correlations are accurately recovered, while the amplitude of HD correlations is underestimated.}
    \label{fig:hdsim_pp_mcos}
\end{figure}

In contrast, the MCOS more accurately recovers the amplitude of monopole and dipole correlations. 
We show the recovered distributions of $\hat{A}^2$ and the S/N 
for models consisting of multiple correlations 
in Figure~\ref{fig:HD_os_hd_mo}.
The distributions for the amplitude and S/N of HD correlations 
are similar to those obtained when fitting for only HD correlations 
(Figure~\ref{fig:HD_os_indv}, top row), 
but now the distributions for the amplitude and S/N of monopole and dipole correlations 
are centered around zero. 
The $p-p$ plots, shown in Figure~\ref{fig:hdsim_pp_mcos}, show the recovery of the MCOS $\hat{A}^2$. We find that MCOS analysis recovers the injected parameters for the monopole and dipole ORFs more accurately than the OS. The amplitude of HD correlations is slightly underestimated, as it was when using the OS.

Table~\ref{tab:hd_sim_results} lists the percentage of ORFs where an S/N greater than 3 was recovered, as well as the percentage of cases where the particular model had the lowest AIC value (most preferential model).
In 51.5 \% of realizations, the standard OS, HD ORF model had the lowest AIC value, as expected, considering our injection has only one signal present. The standard monopole and monopole+HD model have nearly the same preference.
We note that these simulations include relatively low-significance HD correlations, 
as can be seen by the fact that the HD S/N $> 3$ in only 17\% of simulations.
$\bold{B}$ut the percentage of models with an HD S/N $> 3$ does not change drastically between the individual and MCOS models.
This means that it is difficult to distinguish between different correlations. Nevertheless, the MCOS analysis allows us to search for the presence of spatial correlations 
and compute the relative probability of different models.

\subsection{HD + GWMO simulations}

Finally, we looked at how well the MCOS could recover 
multiple correlated signals. 
We generated 200 simulations that contained two stochastic processes: 
one with HD correlations and one with GWMO correlations. 
We used the same amplitude and spectral index $\gamma=13/3$ for both processes. 
We summarize the results of analyzing these simulations in Table~\ref{tab:hd+gwb_sim_results} in Appendix \ref{app:tables}. 

\begin{figure}[t]
        \centering
        \includegraphics[width=\columnwidth]{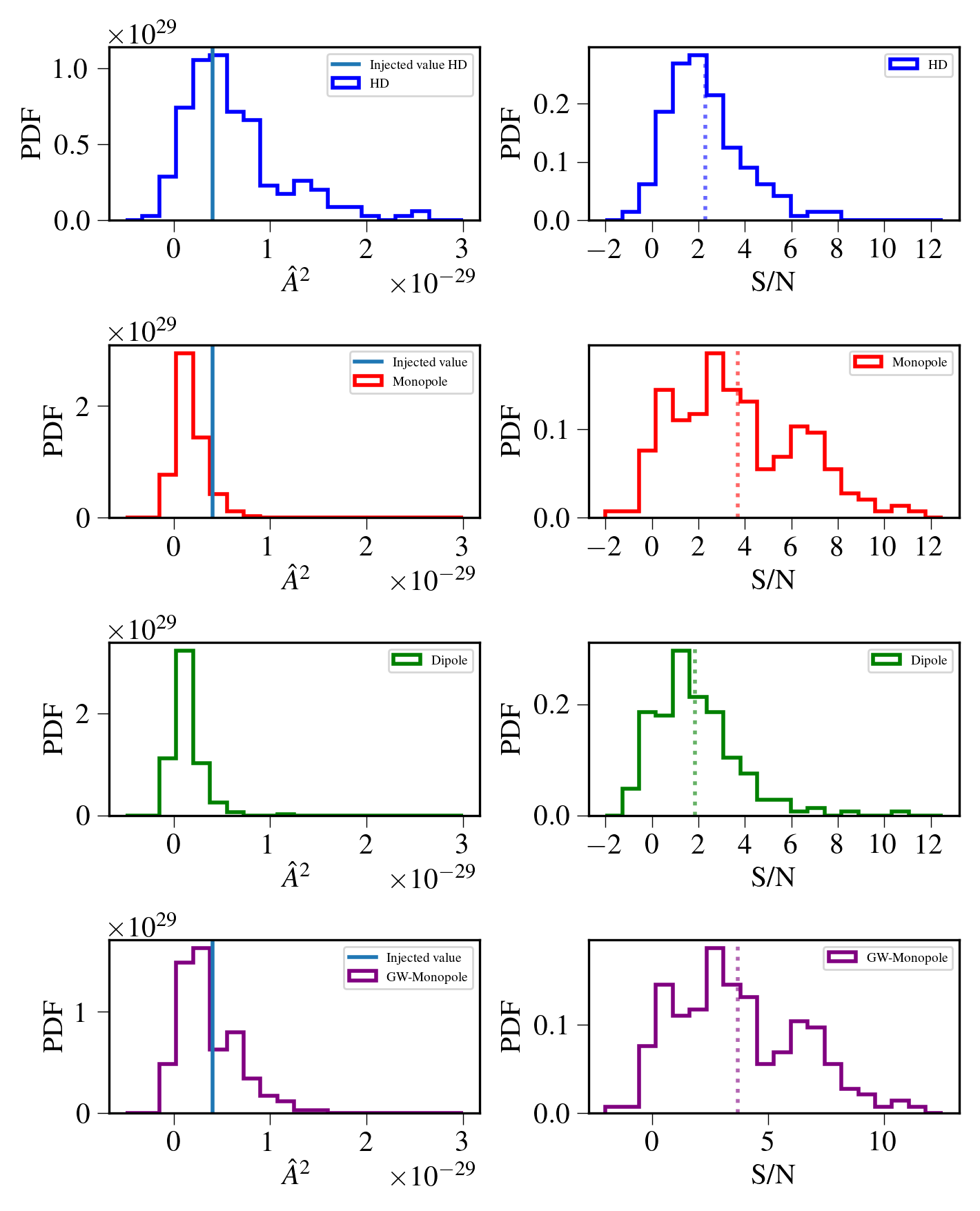}
    \vspace{-0.2in}
        \caption{Individual correlated signal analysis for 200 \textbf{HD}+GWMO simulations. Similar to Fig. \ref{fig:HD_os_indv}, histograms of $\hat{A}^2$ and S/N on the left and right hand columns respectively. From top to bottom, the ORFs are HD (blue), monopole (red), dipole (green), and GWMO (purple). The solid light blue line indicates the injected OS of $A_\mathrm{HD}^2 = 4 \times 10^{-30}$ and $A_\mathrm{GWMO}^2 = 4 \times 10^{-30}$. The colored dashed lines in the S/N plots are
the means of the respective ORFs.}
        \label{fig:gwmo+hd_os_indv}
\end{figure}

When we analyze the simulations for a single correlation, we find S/N $>3$ for HD correlations in 27\% 
of simulations and for GWMO in 56\% of simulations. 
As shown in Figure~\ref{fig:gwmo+hd_os_indv}, we recover the GWMO signal 
with higher mean S/N than the HD signal. 
We note that the correlation coefficient 
for HD correlations is less than the 
correlation coefficient for GWMO correlations for all angular separations (except for zero), 
which is why the OS recovers the GWMO signal with a higher S/N than the HD signal -- 
even though the HD and GWMO signals have the same amplitude, 
the GWMO signal has significantly more power 
in the cross-correlations than the HD signal.

We also recover a dipole signal with S/N $>3$ in 21\% of simulations 
even though no such signal has been injected.

\begin{figure}[t]
        \centering
        \includegraphics[width=0.5\textwidth]{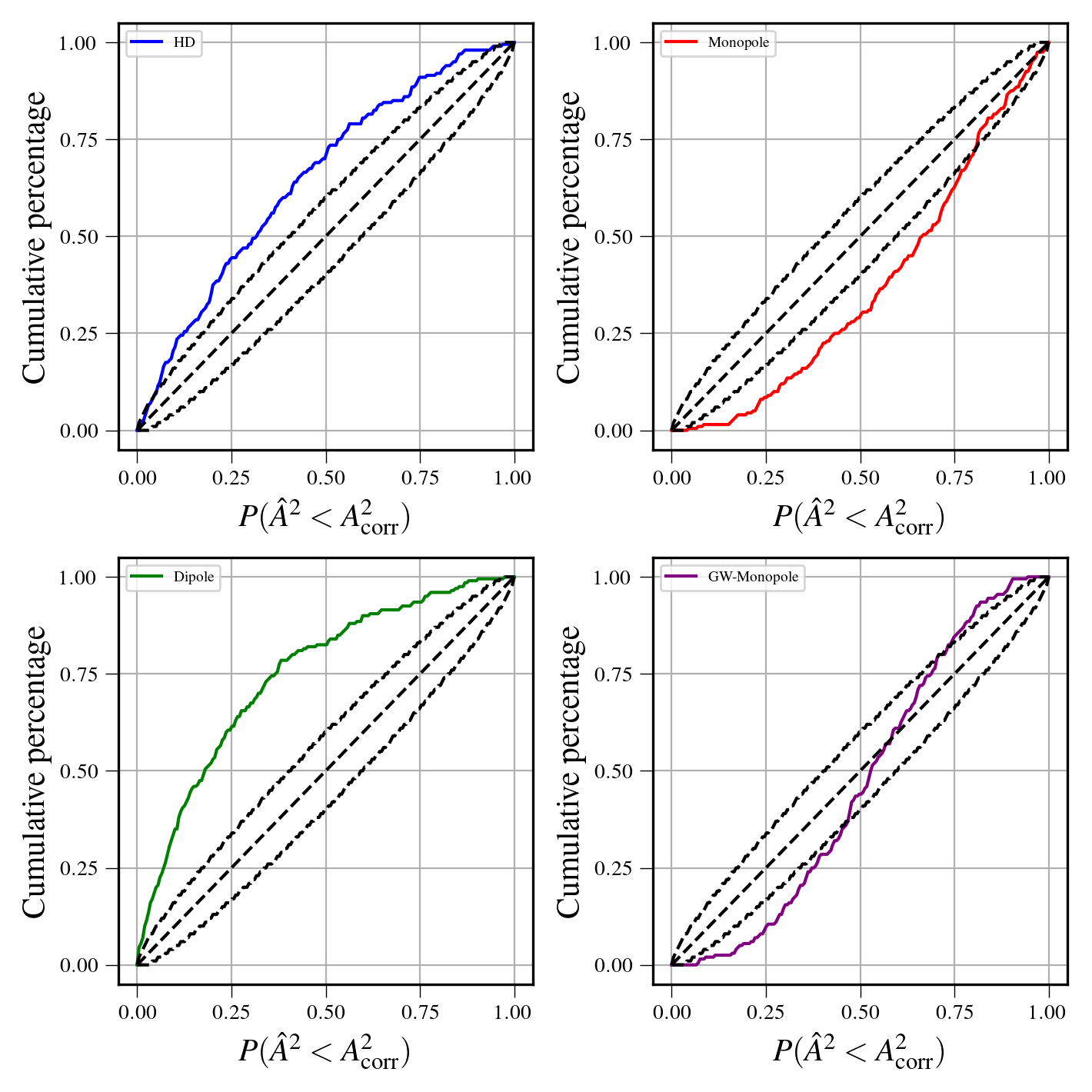}
    \vspace{-0.2in}
        \caption{
        The $p-p$ plots of the HD+GWMO simulations for the individual analysis of (clockwise from top-left) HD, standard monopole, GWMO, and dipole. The horizontal axis is the recovered percentile of each simulation to the injected value, and the vertical axis is the cumulative fraction of realizations that recover the particular percentile. If the percentiles are recovered accurately, the two quantities would be equal (central dashed line). The upper and lower dashed black lines indicate the 95\% confidence interval. Since we are only fitting for one type of correlation, but multiple correlated signals are present in the data, we are not able to accurately recover the amplitudes of the HD or monopole signals. We also overestimate the amplitude of dipole correlations, which are not present.}
        \label{fig:pp_hd+gwmo_ind}
\end{figure}

Figure~\ref{fig:pp_hd+gwmo_ind} shows $\textbf{p-p}$ plots for HD, monopole, dipole, 
and GWMO correlations. 
Since the data contain both HD and GWMO-correlated processes, but we are only fitting 
one process at a time, both signals are being incorrectly associated with a single correlation, 
resulting in inaccurate parameter estimation.
When we fit to HD correlations only, this results in a significant overestimate of the HD-correlated amplitude. 
When we fit to GWMO correlations only, the amplitude estimate is not as affected, 
but the uncertainty is underestimated. 
We also see that the amplitude of a dipolar process is overestimated, even though one is not present.

\begin{figure}[!h]
        \centering
        \includegraphics[width=\columnwidth]{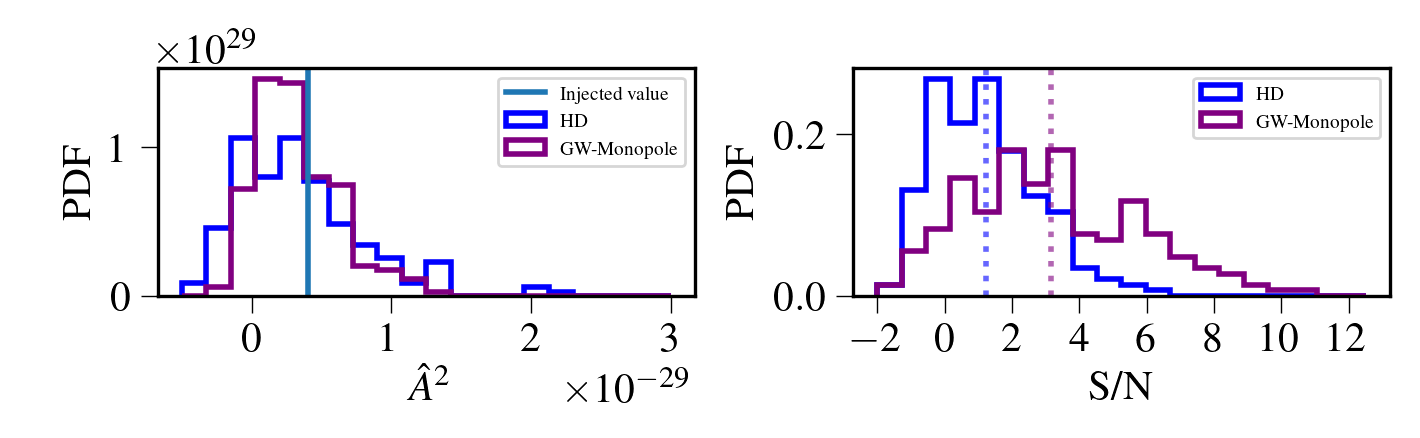}
        \includegraphics[width=\columnwidth]{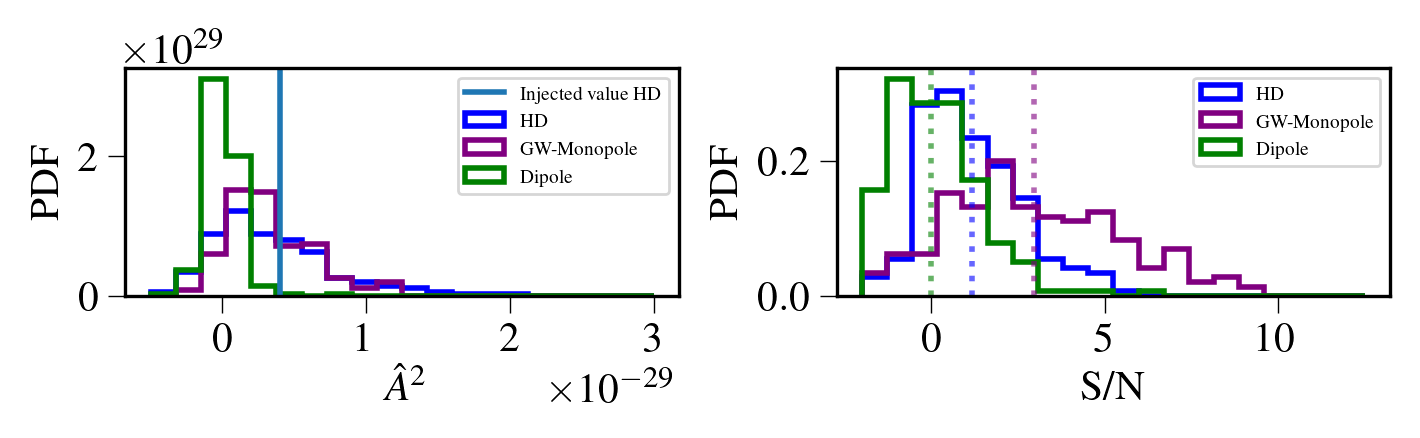}
    \vspace{-0.2in}
        \caption{HD and GWMO multiple correlated signal analysis for 200 HD+GWMO simulations. Similar to Fig. \ref{fig:HD_os_indv}, histograms of the $\hat{A}^2$ and S/N on the left and right hand columns respectively. The solid light blue line indicates the injected OS of $A_\mathrm{HD}^2 = A_\mathrm{GWMO}^2 = 4 \times 10^{-30}$. The colored dashed lines in the S/N plots are the means of the respective ORFs.}
        \label{fig:gwmo+hd_os_hd_gwmo}
\end{figure}

\begin{figure}[t]
        \centering
    \includegraphics[width=\columnwidth]{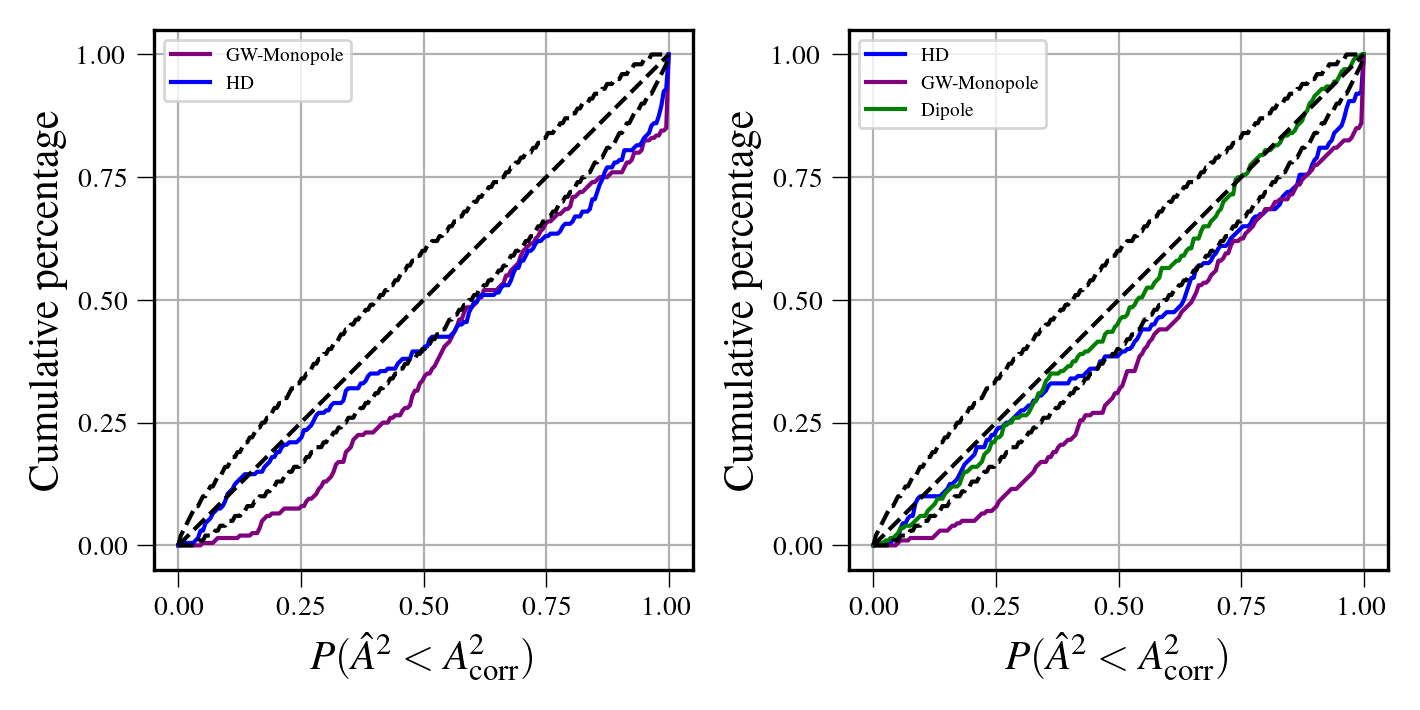}
    \vspace{-0.2in}
        \caption{The $p-p$ plots for the MCOS analysis of the HD+GWMO injections.
        The left figure depicts the HD+GWMO+DI model, and the right figure shows the HD+GWMO model.
        The horizontal axis is the recovered percentile of each simulation to the injected value, and the vertical axis is the cumulative fraction of realizations that recover the particular percentile. The central black dashed line indicates perfectly accurate recovery of percentiles, while the upper and lower black dashed lines indicate the 95\% confidence intervals. We are able to more accurately recover the amplitude of HD and GWMO correlations using the MCOS, although the amplitudes of both are underestimated. We also note that including dipole correlations in the model does not bias parameter estimation of the other signals, and we accurately recover that no dipolar signal is present.}
        \label{fig:pp_hd+gwmo_hd_mo}
\end{figure}

When we use the MCOS, 
we find that we are able to more accurately recover the amplitudes of all the correlated processes. 
Figure~\ref{fig:gwmo+hd_os_hd_gwmo} shows the recovered amplitudes and S/N, 
while Figure~\ref{fig:pp_hd+gwmo_hd_mo} shows $\textbf{p-p}$ plots for the recovered amplitudes. 
When fitting for both HD and GWMO signals, the amplitudes are more accurately recovered 
than when fitting for only one signal at a time, 
although the amplitudes are slightly underestimated. This is possibly 
because of covariance between the pulsar noise and the common signals. 
We also find that when we include dipole correlations in our model, 
we accurately recover that no dipole signal is present, and 
the addition of the dipole signal does not affect the recovery of 
either the HD and GWMO signals.

\section{Conclusions}
\label{sec:conclusions}

In this paper, we present a generalization of the OS
that can simultaneously search for multiple correlated signals. 
As we have shown, this allows us to better distinguish between different spatial correlations. 
For a real PTA, the ORFs corresponding to different correlations are not orthogonal as shown in Table \ref{tab:matchstat}.
The overlap between them can lead to non-zero S/N recovery of non-injected signals. 
It is also useful in cases where multiple correlated signals, e.g., a GWB and correlated noise, are present in the data. 
We have also performed model selection by computing the AIC, 
which acts as a pseudo-Bayes factor.

We have tested this method using two simulated data sets. 
The first contained a stochastic signal with HD correlations. 
When fitting only a single correlation, we found that we overestimated the presence 
of correlations that were not present, 
but using the MCOS eliminated this problem.

We also generated simulated data sets that contained both 
HD and GWMO-correlated signals.
We demonstrated that the MCOS allows both signals to be 
recovered while ruling out any non-injected signals. 

In this paper, we have considered four types of correlations: 
HD, monopole, dipole, and GWMO. 
This method can be used with any number and type of correlations. 
In the NANOGrav 15-year gravitational wave background search \cite{2023ApJ...956L...3A}, 
the MCOS was used to search for evidence of monopolar or dipolar processes, 
which could be associated with correlated noise sources such as clock errors 
or ephemeris errors \cite{2019ApJ...876...55R, 2020ApJ...893..112V, 2020MNRAS.491.5951H}, 
and to do a model-independent reconstruction of the correlations 
using Legendre polynomials \cite{PhysRevD.90.082001}.
It was also used in the search of the NANOGrav 15-year data set for evidence of 
non-Einsteinian polarizations \cite{NANOGrav:2023ygs}. 
The modular nature of the MCOS makes it a valuable tool for characterizing 
the nature of correlations in PTA data sets.

\acknowledgements
We thank Justin Ellis, Jeff Hazboun, David Kaplan, Dusty Madison, Pat Meyers,  Gabe Freedman,
Joe Romano, Xavier Siemens, Abhimanyu Susobhanan, and Michele Vallisneri 
for useful discussions. 
We thank Nima Laal for help creating simulated data sets 
with monopole injections.

This work was supported by 
National Science Foundation (NSF) grant PHY-2011772. SRT acknowledges support from NSF AST-2007993 and AST-2307719. SRT and KAG also acknowledge support from an NSF CAREER \#2146016.
The authors are members of the NANOGrav collaboration, 
which receives support from NSF Physics Frontiers Center 
award number 1430284 and 2020265.

\newpage

\appendix

\section{Simulation results}
\label{app:tables}

This appendix contains the results of the MCOS analysis on the HD (Table \ref{tab:hd_sim_results}), and HD+GWMO  (Table \ref{tab:hd+gwb_sim_results}) simulations. The middle three columns refer to the percentage of realizations where the ORF listed in the top row had a S/N greater than or equal to three. The last column shows the percentage of realizations where the particular model listed was the most favored, with a $p(AIC) = 1$. Since certain realizations had AIC values very similar to one another, implying that more than one model was equally favored, we have set our p(AIC) threshold to 0.99.

\vspace{-1in}%
\input{table_hd_results}%
\vspace{-1in}%
\input{table_hdgwmo_results}%

\clearpage

\section{Pulsar Red Noise Parameters}
\label{app:pulsar_noise}

Our simulated data sets include intrinsic red noise 
in each pulsar. We model the red noise with a power-law power spectrum 
$P(f) = \frac{A^2}{12 \pi^2} \left(\frac{f}{f_\mathrm{yr}}\right)^{-\gamma} \mathrm{yr}^3 $ (as shown in Equation \ref{eq:rn_psd}. 
The values of $A$ and $\gamma$ used for each pulsar are listed in 
Table~\ref{tab:rednoise}, 
and are based on the noise properties of the NANOGrav 12.5-year data set.  

\input{table_psr_params}

\section{Incorporating Pair Covariance into the OS and MCOS}
\label{app:p-p}

In the weak-signal regime, 
our pulsar pair cross-correlations $\rho_{ab}$ and their associated uncertainties $\sigma_{ab}$ as described in Equations \ref{eq:rho_ab} and \ref{eq:sig_ab}, are given by \cite{2013ApJ...762...94D,2022ApJ...940..173P,2015PhRvD..91d4048C,2013CQGra..30v4015S,2018PhRvD..98d4003V}

\begin{eqnarray}
    \rho_{ab} &=& \mathcal{N}_{ab} \mathbf{r^{T}_{a} P^{-1}_{a} \Bar{S}_{ab} P^{-1}_{b} r_{b}} \equiv \mathbf{r_{a}^{T} Q_{ab} r_{b}} \,, \\ \label{eq:rho1}
    \sigma_{ab}^2 &=& \mathcal{N}_{ab} = (\mathrm{tr}[\mathbf{P^{-1}_{a} \Bar{S}_{ab} P^{-1}_{b} \Bar{S}_{ba}} ])^{-1} \,, \label{eq:sig1}
\end{eqnarray}
where $\mathbf{P}_{a}$ is the auto-covariance matrix of pulsar $a$, $\mathbf{S}_{ab} = A_\mathrm{gw}^2 \Gamma_{ab} \mathbf{\Bar{S}}_{ab}$ is the cross-covariance matrix between pulsars $a$ and $b$, 
and $\mathbf{Q_{ab}} \equiv \mathcal{N}_{ab} \mathbf{P_{a}^{-1} \Bar{S}_{ab} P_{b}^{-1}}$. The auto-correlation matrices $\mathbf{P}_a$ contain contributions from white noise, intrinsic red noise, and the common process red noise, while the cross-correlation matrices $\mathbf{S}_{ab}$ only considers contributions from the common process. 
The definition of the cross-correlation uncertainties 
$\sigma_{ab}$ in Equation~\ref{eq:sig1} is only valid in the 
weak-signal regime and does not include correlations 
between different pairs of pulsars arising 
from the fact that some of the pulsar pairs will have pulsars 
in common (e.g., the correlations $\rho_{ab}$ and $\rho_{ac}$ both have the pulsar $a$ in common).

If a significant correlated signal is present, 
then $\sigma_{ab}$ is an underestimate of the cross-correlation uncertainties. 
When we consider the covariances between different pairs of pulsars, our uncertainties $\sigma_{ab}$ must include an extra term 
for inter-pulsar pair covariances. 
For two pulsar pairs $ab$ and $cd$ 
\cite{2023arXiv230616223J},

\begin{eqnarray}
    \mathbf{\Sigma}_{ab,cd} &=& \left\langle \rho_{ab} \rho_{cd} \right\rangle - \left\langle \rho_{ab} \right\rangle \left\langle \rho_{cd} \right\rangle \\
    &=& \mathrm{tr}\left[ \mathbf{Q_{ba} S_{ac} Q_{cd} S_{db}} \right] \nonumber \\
    && + \mathrm{tr}\left[ \mathbf{Q_{ba} S_{ad} Q_{dc} S_{cb}} \right] \,.
\end{eqnarray}

\noindent Note that $\mathbf{\Sigma}_{ab,cd}$ 
is dependent on the ORF and amplitude of the common process, due to the contributions from $\mathbf{S}_{\mathrm{ab}}$.

For the standard single component OS,

\begin{eqnarray}
    \mathbf{\Sigma^\mathrm{OS}}_{ab,cd} 
    &=& A_{\mathrm{CURN}}^4 \Gamma_{ac} \Gamma_{db} \mathrm{tr}\left[ \mathbf{Q_{ba} \Bar{S}_{ac} Q_{cd} \Bar{S}_{db}} \right] \nonumber \\
    &&+ A_{\mathrm{CURN}}^4 \Gamma_{ad} \Gamma_{cb} \mathrm{tr}\left[ \mathbf{Q_{ba} \Bar{S}_{ad} Q_{dc} \Bar{S}_{cb}} \right] \,,
\end{eqnarray}
where $\Gamma$ is the ORF for the correlation, and $A_\mathrm{CURN}^2$ is the amplitude of the common uncorrelated red noise found from a Bayesian analysis. 
We chose to use the amplitude of the common uncorrelated red noise here instead of the amplitude of the correlated process because it .

When we include pair covariance, the expressions for 
the OS and its variance become
\begin{eqnarray}
    \hat{A}^2 &=& \frac{\sum_{ab,cd}\Gamma_{ab}\left(\mathbf{\Sigma}^\mathrm{OS}_{ab,cd}\right)^{-1}\rho_{cd}}{\sum_{ab,cd}\Gamma_{ab}\left(\mathbf{\Sigma}^{\mathrm{OS}}_{ab,cd} \right)^{-1}\Gamma_{cd}} \,, \label{eq:os_paircov} \\
    \sigma_{\hat{A}^2} &=& \left[ \sum_{ab,cd} \Gamma_{ab} \left(\mathbf{\Sigma}^\mathrm{OS}_{ab,cd}\right)^{-1} \Gamma_{cd} \right]^{-\frac{1}{2}} \,. \label{eq:oserr_paircov}
\end{eqnarray}

For the MCOS, we generalize Eqs.~\eqref{eq:os_paircov} and \eqref{eq:oserr_paircov} to allow for arrays of $A^2$ and $\Gamma$:
\begin{equation}
    \mathbf{S}_{ab} = \mathbf{\Bar{S}}_{ab} \Gamma_{ab} A^{2} = \mathbf{\Bar{S}}_{ab} \sum_{i} \Gamma_{ab}^{i} A^{2}_{i} \,,
\end{equation}
\noindent where $i$ is the index of the ORF.

To obtain the amplitudes of the correlated signals $A_i^2$, 
we first calculate the MCOS values neglecting pair covariance 
using Eq.~\eqref{eq:OS_MCOS_A2}. 

If any of the values for $\hat{A}^2$ are negative, we set it equal to zero. We then set $A^2_i$ to be equal to a weighted fraction of the common uncorrelated red noise amplitude,
\begin{eqnarray}
    A^2_i = A^2_\mathrm{CURN} \times \frac{A^2_{\mathrm{MCOS}, i}}{\sum_{j} A^2_{\mathrm{MCOS}, j}} \,.
\end{eqnarray}
Then the MCOS covariance matrix becomes

\begin{eqnarray}
    \mathbf{\Sigma}_{ab,cd}^{\mathrm{MCOS}} =&& (\sum_{i} A^2_{i} \Gamma_{ac}^{i}) (\sum_{j} A^2_{j} \Gamma_{db}^{j}) \mathrm{tr}\left[ \mathbf{Q_{ba} \Bar{S}_{ac} Q_{cd} \Bar{S}_{db}} \right] \nonumber \\
    &+& (\sum_{i} A^2_{i} \Gamma_{ad}^{i}) (\sum_{j} A^2_{j} \Gamma_{cb}^{j}) \nonumber \\
    &\times& \mathrm{tr}\left[ \mathbf{Q_{ba} \Bar{S}_{ad} Q_{dc} \Bar{S}_{cb}} \right] \,,
\end{eqnarray}

\noindent then, as we did in Section \ref{sec:os}, we can replace the $\sigma_{ab}^{2}$ with our new covariance $\mathbf{\Sigma}^{\mathrm{MCOS}}_{ab,cd}$ and construct our new MCOS $B$ and $C$ matrices

\begin{eqnarray}
    C^\beta &\equiv& \sum_{ab,cd} \rho_{ab} \left[ \mathbf{\Sigma}^{\mathrm{MCOS}}_{ab,cd} \right]^{-1} \Gamma_{cd}^\beta \,, \\
    B^{\alpha\beta} &\equiv& \sum_{ab,cd} \Gamma_{ab}^\alpha \left[ \mathbf{\Sigma}^{\mathrm{MCOS}}_{ab,cd} \right]^{-1} \Gamma_{cd}^\beta \,, \\
    \hat{A}^2_\alpha &=& \sum_\beta B_{\alpha\beta} C^\beta \,, \\
    \sigma_{\hat{A}^2_\alpha}^2 &=& B_{\alpha\alpha} \,.
\end{eqnarray}

\bibliographystyle{apsrev}
\bibliography{authors}

\end{document}

%% file: table_orfs.tex
\begin{table}
    \begin{center}
    \caption{Match statistic $\bar{M}$ for HD, monopole, dipole, and GW-like monopole correlations, as defined by Eq.~\eqref{eq:orf_match}, for our simulated PTA. The monopole and GW-like monopole have $\bar{M} = 1$ 
since they only differ by a normalization factor.}
    \label{tab:matchstat}
    \begin{tabular}{ l|cccc } 
    \hline \hline
         & \hspace{0.5in} &  &  & GW-like\\
        Correlation & HD & Monopole & Dipole & monopole\\
    \hline
        HD & 1 & 0.255 & 0.435 & 0.255\\ 
        Monopole & $\cdots$ & 1 & 0.395 & 1\\ 
        Dipole & $\cdots$ & $\cdots$ & 1 & 0.395\\
        GW-like & \multirow{2}{*}{$\cdots$} & \multirow{2}{*}{$\cdots$} & \multirow{2}{*}{$\cdots$} & \multirow{2}{*}{1} \\
        monopole &  &  &  &  \\
    \hline
    \end{tabular}
    \end{center}
\end{table}

%% file: table_hd_results.tex
\begin{table*}[t]
    \begin{center}
    \caption{Summary of results of GWB simulations.}
    \label{tab:hd_sim_results}
    \begin{tabular}{l|ccc|c}
    \hline \hline
        & \multicolumn{3}{c|}{\% of Simulations with S/N $>3$} & \% of Simulations \\
        Model & HD & Monopole & Dipole & with $p(\mathrm{AIC}) \geq 0.99$\\
    \hline
        HD only & 17.5\% & $\cdots$ & $\cdots$ & 40\% \\ 
        Monopole only & $\cdots$ & 7\% & $\cdots$ & 18\%\\ 
        Dipole only & $\cdots$ & $\cdots$ & 5\% & 14\%\\
        HD + monopole & 16.5\% & 2\% & $\cdots$ & 15.5\%\\
        HD + dipole & 16\% & $\cdots$ & 1\% & 8.5\%\\
        Monopole + dipole & $\cdots$ & 3\% & 2\% & 1\%\\
        HD + monopole + dipole & 14.5\% & 2\% & 0.5\% & 5\%\\
    \hline
    \end{tabular}
    \end{center}
\end{table*}

%% file: table_hdgwmo_results.tex
\begin{table*}
    \begin{center}
    \caption{Summary of results of simulations with both GWB and GWMO stochastic processes.}
    \label{tab:hd+gwb_sim_results}
    \begin{tabular}{l|cccc|c} 
       \hline \hline
        & \multicolumn{4}{c|}{\% of Simulations with S/N $>3$} & \% of Simulations \\
            Model & HD  & GWMO & DI & MO & with $p(\mathrm{AIC}) \geq 0.99$\\
        \hline
            HD & 27.5\% & $\cdots$ & $\cdots$ & $\cdots$ & 13.5\%\\ 
            GWMO & $\cdots$ & 53.5\% & $\cdots$ & $\cdots$ & 39\%\\ 
            Dipole & $\cdots$ & $\cdots$ & 21\% & $\cdots$ & 7.5\%\\
            Monopole & $\cdots$ & $\cdots$ & $\cdots$ & 53.5\% & 39\% \\
            HD+GWMO+Dipole & 10.5\% & 44\% & 2\% & $\cdots$ & 18.5\%\\
            HD+Monopole+Dipole & 10.5\% & $\cdots$ & 2\% & 44\% & 18.5\%\\
            HD+GWMO & 13.5\% & 48\% & $\cdots$ & $\cdots$ & 19.5\%\\
            HD+Monopole & 13.5\% & $\cdots$ & $\cdots$ & 48\% & 19.5\%\\
            HD+Dipole & 16.5\% & $\cdots$ & 10\% & $\cdots$ & 2.5\%\\
            GWMO+Dipole & $\cdots$ & 46.5\% & 5\% & $\cdots$ & 0\%\\
        \hline
    \end{tabular}
    \end{center}
\end{table*}

%% file: table_psr_params.tex
\begin{table}[h!]
    \centering
    \caption{Pulsar red noise parameters $A$ and $\gamma$ 
    used in our simulated data sets.   
    }
    \label{tab:rednoise}
    \begin{adjustbox}{max width=8cm}
    \begin{tabular}{l|cc}
    \hline \hline
    Pulsar & $\log_{10}A$ injected & $\gamma$ injected\\
    \hline
    B1855+09 & -14.3053 & 5.5879 \\
    B1937+21 & -13.4731 & 3.6189 \\
    B1953+29 & -12.7813 & 1.1485 \\
    J0023+0923 & -13.2146& 0.2572 \\
    J0030+0451 & -14.8837 & 5.1835 \\
    J0340+4130 & -17.2031 & 0.3121 \\
    J0613$-$0200 & -17.8287 & 6.9765 \\
    J0636+5128 & -16.6117 & 0.4274 \\
    J0645+5158 & -19.3925 & 6.2542 \\
    J0740+6620 & -18.3882 & 0.8377 \\
    J0931$-$1902 & -19.3076 & 3.1884 \\
    J1012+5307 & -12.7970 & 0.9968 \\
    J1024$-$0719 & -17.2779 & 2.5507 \\
    J1125+7819 & -12.4459 & 1.5990 \\
    J1453+1902 & -18.9550 & 4.9088 \\
    J1455$-$3330 & -19.2175 & 6.2343 \\
    J1600$-$3053 & -13.3480 & 0.0627 \\
    J1614$-$2230 & -19.8669 & 2.5180 \\
    J1640+2224 & -16.3220 & 5.6796 \\
    J1643$-$1224 & -12.2480 & 1.1554 \\
    J1713+0747 & -14.0354 & 0.3368 \\
    J1738+0333 & -19.5594 & 5.2048 \\
    J1741+1351 & -14.3474 & 1.9857 \\
    J1744$-$1134 & -13.3903 & 1.5453 \\
    J1747$-$4036 & -12.7570 & 3.5500 \\
    J1832$-$0836 & -13.1021 & 1.6074 \\
    J1853+1303 & -13.3632 & 1.3637 \\
    J1903+0327 & -12.2570 & 1.8621 \\
    J1909$-$3744 & -13.9843 & 1.7386 \\
    J1910+1256 & -16.0932 & 2.6105 \\
    J1911+1347 & -13.4836 & 1.7034 \\
    J1918$-$0642 & -19.7173 & 4.0390 \\
    J1923+2515 & -17.7267 & 1.7844 \\
    J1944+0907 & -13.2678 & 1.2939 \\
    J2010$-$1323 & -18.9887 & 0.8175 \\
    J2017+0603 & -17.8317 & 1.9894 \\
    J2033+1734 & -17.3183 & 3.8762 \\
    J2043+1711 & -17.5979 & 2.8240 \\
    J2145$-$0750 & -12.9290 & 1.2728 \\
    J2214+3000 & -16.9390 & 4.6045 \\
    J2229+2643 &-17.4349 & 6.1385\\
    J2234+0611 & -13.5827 & 3.6588 \\
    J2234+0944 & -18.1712 & 0.7777 \\
    J2302+4442 & -14.7765 & 2.8715 \\
    J2317+1439 & -18.1829 & 6.5325 \\
    \hline
    \end{tabular}
\end{adjustbox}
\end{table}